\documentclass[floatfix,twocolumn,showpacs,preprintnumbers,amsmath,amssymb,pra,superscriptaddress,longbibliography]{revtex4-1}
\usepackage{color}
\usepackage[usenames,dvipsnames,svgnames,table]{xcolor}
\usepackage[colorlinks=true,linkcolor=blue,urlcolor=blue,citecolor=blue]{hyperref}
\usepackage{mathtools}
\usepackage{graphicx}
\usepackage{dcolumn}
\usepackage{array}
\usepackage{lipsum}
\usepackage{bm}
\usepackage{subfigure}
\usepackage{amssymb}
\usepackage{multirow}
\usepackage{tabularx}
\usepackage{amsmath}
\usepackage{braket}
\graphicspath{{plots/}}
 \usepackage{lipsum}
\usepackage{mathrsfs}

\newcommand{\beq}{\begin{equation}}
\newcommand{\eeq}{\end{equation}}
\newcommand{\bea}{\begin{eqnarray}}
\newcommand{\eea}{\end{eqnarray}}

\newcommand{\qv}{{\bf q}}

\renewcommand{\vec}[1]{\mathbf{#1}}

\begin{document}
\title{
Nonlinear electronic density response of the warm dense electron gas: multiple perturbations and mode coupling
}

\author{Tobias Dornheim}
\email{t.dornheim@hzdr.de}

\affiliation{Center for Advanced Systems Understanding (CASUS), D-02826 G\"orlitz, Germany}
\affiliation{Helmholtz-Zentrum Dresden-Rossendorf (HZDR), D-01328 Dresden, Germany}

\author{Jan Vorberger}
\affiliation{Helmholtz-Zentrum Dresden-Rossendorf (HZDR), D-01328 Dresden, Germany}

\author{Zhandos A.~Moldabekov}

\affiliation{Center for Advanced Systems Understanding (CASUS), D-02826 G\"orlitz, Germany}
\affiliation{Helmholtz-Zentrum Dresden-Rossendorf (HZDR), D-01328 Dresden, Germany}

\author{Michael Bonitz}

\affiliation{Institut f\"ur Theoretische Physik und Astrophysik, Christian-Albrechts-Universit\"at zu Kiel, D-24098 Kiel, Germany}

\begin{abstract}
We present extensive new \emph{ab initio} path integral Monte Carlo (PIMC) results for an electron gas at warm dense matter conditions that is subject to multiple harmonic perturbations. In addition to the previously investigated nonlinear effects at the original wave number [Dornheim \emph{et al.}, PRL \textbf{125}, 085001 (2020)] and the excitation of higher harmonics [Dornheim \emph{et al.}, PRR \textbf{3}, 033231 (2021)], the presence of multiple external potentials leads to mode-coupling effects, which constitute the dominant nonlinear effect and lead to a substantially more complicated density response compared to linear response theory. One possibility to estimate mode-coupling effects from a PIMC simulation of the unperturbed system is given in terms of generalized imaginary-time correlation functions that have been recently introduced by Dornheim \emph{et al.}~[JCP \textbf{155}, 054110 (2021)]. In addition, we extend our previous analytical theory of the nonlinear density response of the electron gas in terms of the static local field correction [Dornheim \emph{et al.}, PRL \textbf{125}, 235001 (2020)], which allows for a highly accurate description of the PIMC results with negligible computational cost.
\end{abstract}

\maketitle

\section{Introduction\label{sec:introduction}}

The uniform electron gas (UEG) is one of the most fundamental model systems in physics and quantum chemistry~\cite{loos,quantum_theory} and has facilitated important insights into a gamut of physical effects such as Fermi liquid theory~\cite{quantum_theory} and the Bardeen-Cooper-Schrieffer theory of superconductivity~\cite{Bardeen_PhysRev_1957}. Moreover, the accurate description of the UEG based on quantum Monte Carlo (QMC) simulations in the electronic ground state~\cite{Ceperley_Alder_PRL_1980,moroni2,Spink_PRB_2013} has been pivotal for many practical applications such as density functional theory (DFT) simulations of real materials~\cite{Jones_RMP_2015}.

Over the last decades, there has emerged a growing interest into the properties of the UEG at extreme temperatures and densities~\cite{review,fortov_review}. Such \emph{warm dense matter} (WDM) naturally occurs in astrophysical objects like giant planet interiors~\cite{militzer1,Militzer_2008} and brown dwarfs~\cite{becker,saumon1}. In addition, it has been predicted to occur on the pathway towards inertial confinement fusion~\cite{hu_ICF} and might be used as a catalyst to accelerate chemical reactions~\cite{Brongersma2015}. Consequently, WDM is nowadays routinely realized in the laboratory using different experimental techniques; see Ref.~\cite{falk_wdm} for a recent review article. 

This interest in the properties of WDM has led to a spark of new developments in the QMC simulation of the UEG at finite temperature~\cite{Brown_PRL_2013,Malone_PRL_2016,dornheim_prl,dornheim_pop17,dornheim_jcp15,dornheim_prb16,groth_prb16,schoof_prl15,Joonho_JCP_2021}, which culminated in the first accurate parametrization of the exchange--correlation (XC) free energy $f_\textnormal{xc}$ covering the entire relevant range of densities and temperatures~\cite{groth_prl,ksdt,Karasiev_status_2019,review}.
These new XC-functionals have already been utilized for thermal DFT simulations of WDM~\cite{karasiev_importance,kushal}, which has further substantiated the impact of thermal effects on material properties for some parameters. 

In addition to $f_\textnormal{xc}$, accurate data for the warm dense UEG have been presented for a multitude of quantities including the static structure factor $S(\mathbf{k})$~\cite{dornheim_cpp17,Joonho_JCP_2021,Dornheim_PRL_2020,Dornheim_PRL_2020_ESA}, the momentum distribution function $n(\mathbf{k})$~\cite{MILITZER201913,hunger_20,Dornheim_PRB_nk_2021,dornheim2021momentum,Militzer_Pollock_PRL_2002} the dynamic structure factor $S(\mathbf{k},\omega)$~\cite{dornheim_dynamic,dynamic_folgepaper,Dornheim_PRE_2020}, and other dynamic properties like the conductivity~\cite{hamann_prb_20,hamann_cpp_20}.

Of particular importance for many practical applications~\cite{kraus_xrts,siegfried_review,ceperley_potential,moldabekov_pre_20} in WDM theory is the response of the UEG to an external perturbation~\cite{quantum_theory}. Typically, such effects are described within linear response theory (LRT), which presupposes a simple, linear relation between response and perturbation and, thus, leads to a drastic simplification of the underlying theory. Therefore, accurate data for the linear response of the UEG based on QMC simulations are available both in the electronic ground state~\cite{moroni,moroni2,cdop} and at finite temperature~\cite{dornheim_ML,dornheim_electron_liquid,dornheim_HEDP,castello2021classical,Dornheim_PRB_ESA_2021}.

Very recently, Dornheim \textit{et al.}~\cite{dornheim_prl_20,Dornheim_PRR_2021,Dornheim_CPP_2021} have gone beyond the assumption of LRT by carrying out extensive \emph{ab initio} path integral Monte Carlo (PIMC) simulations of a harmonically perturbed electron gas~\cite{moroni,dornheim_pre17,groth_jcp17,bowen2}. First and foremost, this has allowed for the first time to unambiguously check the validity range of LRT. Indeed, it has been reported that nonlinear effects can be important in some situations that are of relevance for state-of-the-art WDM experiments, e.g. using free electron lasers~\cite{Fletcher2015}.
In addition, these PIMC studies have allowed to obtain the first data for various nonlinear density response functions both at the wave vector of the original perturbation~\cite{dornheim_prl_20} and its integer harmonics~\cite{Dornheim_PRR_2021}. This, in turn, has allowed the same group to present an accurate analytical theory for the description of the nonlinear electronic density response in terms of the effectively static local field correction (LFC) $\overline{G}(\mathbf{k})$, which is available as a readily usable analytical representation~\cite{Dornheim_PRB_ESA_2021}.

Further new results into this direction include the exploration of the straightforward relation between nonlinear effects and higher order correlation functions known from many-body theory~\cite{Dornheim_JPSJ_2021} and the computation of the nonlinear density response based on imaginary-time correlation functions defined with respect of the unperturbed system~\cite{Dornheim_JCP_ITCF_2021}.

Yet, all of these works have been restricted to the study of nonlinear effects that emerge as the consequence of a single harmonic perturbation. This constitutes a serious restriction, as realistic experimental perturbations are often given as a superposition of many harmonics~\cite{siegfried_review,Hartmann2018}. 
In fact, there exist are large manifold of experimental techniques involving combinations of two or more frequencies. Examples are four wave mixing or Raman spectroscopy, including Stokes and Anti-Stokes Raman scattering, stimulated Raman spectroscopy (SRS) or coherent anti-Stokes Raman spectroscopy (CARS). Also, the plasmon signal in XRTS of WDM can be viewed as a combination of X-ray and plasmon oscillations.
Naturally, this is of no consequence within LRT, where the response to a superposition of perturbations is simply given by the superposition of the individual responses. This drastically changes once nonlinear effects are taken into account, which leads to interesting and non-trivial mode-coupling effects.

In the present work, we rigorously study these mode-coupling effects by carrying out extensive new PIMC simulations of a warm dense electron gas that is subject to multiple harmonic perturbations. Remarkably, we find that mode-coupling constitutes the dominant nonlinear effect for weak to moderate values of the perturbation amplitude $A$. In addition, we practically demonstrate that these effects can be estimated from the imaginary-time structure of the unperturbed system. Finally, we extend our earlier analytical theory of the nonlinear density response of the UEG~\cite{Dornheim_PRR_2021} and find excellent agreement between our LFC-based theory and the exact PIMC results.

For completeness, we note that the mode coupling effects observed in our paper should not be confused with mode coupling theory (MCT). MCT means linear superposition of field modes, and this term is used extensively in electromagnetism, nanophotonics and other fields which are based on Maxwell's equations the linearity of which implies the superposition principle of individual modes. Deviations occur only in the case that the field penetrates a nonlinear medium. This is, in fact, the case of WDM.

The paper is organized as follows: In Sec.~\ref{sec:theory}, we introduce the relevant theoretical background including the model system (\ref{sec:system}) and different approaches to the density response (\ref{sec:response}). Sec.~\ref{sec:results} is devoted to the discussion of our new simulation results, and the paper is concluded by a brief summary and outlook in Sec.~\ref{sec:summary}.

\section{Theory\label{sec:theory}}
We assume Hartree atomic units throughout this work. Furthermore, we restrict ourselves to a fully unpolarized (\emph{paramagnetic}) system where the number of spin-up and -down electrons are equal, i.e., $N^\uparrow=N^\downarrow=N/2$.

\subsection{Model system\label{sec:system}}

The uniform electron gas is typically characterized by two dimensionless parameters, which are both of the order of unity in the WDM regime~\cite{ott_epjd18}: a) the density parameter (also known as Wigner-Seitz radius) $r_s=\overline{r}/a_\textnormal{B}$, where $\overline{r}$ and $a_\textnormal{B}$ are the average inter-particle distance and first Bohr radius, and b) the degeneracy temperature $\theta=k_\textnormal{B}T/E_\textnormal{F}$, with $E_\textnormal{F}=k_\textnormal{F}^2/2$, and the Fermi wave number being defined as
\begin{eqnarray}
k_\textnormal{F} = \left(
\frac{9\pi}{4}
\right)^{1/3} r_s^{-1}\ .
\end{eqnarray}
From a physical perspective, these two parameters allow for a straight forward interpretation: for $\theta\gg1$, quantum effects are negligible and the system attains the classical limit, whereas $\theta\ll1$ indicates the electronic ground state, where  quantum degeneracy effects predominate; regarding the Wigner-Seitz radius, $r_s\ll1$ indicates the weak coupling regime, and the UEG actually converges towards the ideal Fermi gas in the limit $r_s\to0$. In contrast, $r_s\gg1$ indicates strong coupling, where the UEG first transforms into an electron liquid~\cite{dornheim_electron_liquid,castello2021classical} and subsequently forms a Wigner crystal~\cite{PhysRevB.69.085116}. Thus, the density parameter plays the role of an effective coupling parameter for the UEG, which can be easily seen by considering the kinetic and interaction contributions $K$ and $W$ in their respective lowest order: the Hartree-Fock kinetic energy is given by $K=a_\textnormal{HF}/r_s^2$, and the mean-field part of the interaction scales as $W=b/r_s$. The usual definition of a coupling parameter thus directly results in $\Gamma=W/K\sim r_s$.

Let us consider a Hamiltonian of the form
\begin{eqnarray}\label{eq:Hamiltonian}
\hat H = \hat H_\textnormal{UEG} + \sum_{l=1}^N V(\hat{\mathbf{r}}_l)\ ,
\end{eqnarray}
where $\hat H_\textnormal{UEG}$ is the standard Hamiltonian of the unperturbed UEG~\cite{review} and the external potential $V(\mathbf{r})$ is comprised of multiple harmonic perturbations,
\beq
V(\mathbf{r})=\sum_i A_i\cos(\qv_i\mathbf{r})\,,
\eeq
which immediately leads to the Fourier transform
\beq
 V(\qv)=\sum_i A_i\left[\delta(\qv-\qv_i)+\delta(\qv+\qv_i)\right]\,.
\eeq

In practice, we simulate the inhomogeneous electron gas that is governed by Eq.~(\ref{eq:Hamiltonian}) using the direct PIMC method~\cite{cep,dornheim_jcp_19} without any nodal constraints. Therefore, our simulations are computationally involved due to the fermion sign problem~\cite{troyer,dornheim_sign_problem,Dornheim_2021}, but exact within the given statistical uncertainty.

\subsection{Density response\label{sec:response}}

The induced density contains then a variety of combinations of higher harmonics of the incoming perturbing potential
\begin{widetext}
\bea\label{eq:response}
n_{ind}(\qv)&=&\chi(\mathbf{q}) \sum_i A_i\Big\{ \delta(\qv-\qv_i)+\delta(\qv+\qv_i)\Big\}  \nonumber\\
&+&\sum_{ij}A_iA_j\Big\{ 
Y(\qv-\qv_i,\qv_i)\left[\delta(\qv-\qv_i-\qv_j)+\delta(\qv-\qv_i+\qv_j)\right]\nonumber\\
&&\qquad\qquad+Y(\qv+\qv_i,-\qv_i)\left[\delta(\qv+\qv_i-\qv_j)+\delta(\qv+\qv_i+\qv_j)\right]
\Big\}\nonumber\\
&+&\sum_{ijl}A_iA_jA_l\Big\{
Z(\qv-\qv_j-\qv_l,\qv_j,\qv_l)\left[\delta(\qv-\qv_i-\qv_j-\qv_l)+\delta(\qv+\qv_i-\qv_j-\qv_l)\right]
\nonumber\\
&&\qquad\qquad\quad
+Z(\qv+\qv_j-\qv_l,-\qv_j,\qv_l)\left[\delta(\qv-\qv_i+\qv_j-\qv_l)+\delta(\qv+\qv_i+\qv_j-\qv_l)\right]\nonumber\\
&&\qquad\qquad\quad
+Z(\qv-\qv_j+\qv_l,\qv_j,-\qv_l)\left[\delta(\qv-\qv_i-\qv_j+\qv_l)+\delta(\qv+\qv_i-\qv_j+\qv_l)\right]\nonumber\\
&&\qquad\qquad\quad
+Z(\qv+\qv_j+\qv_l,-\qv_j,-\qv_l)\left[\delta(\qv-\qv_i+\qv_j+\qv_l)+\delta(\qv+\qv_i+\qv_j+\qv_l)\right]
\Big\}\,.
\eea
\end{widetext}
Specifically, $\chi(\mathbf{q})$ denotes the usual static limit of the linear response function~\cite{quantum_theory,nolting}, $Y(\mathbf{q}_1,\mathbf{q}_2)$ the generalized quadratic response function, and $Z(\mathbf{q}_1,\mathbf{q}_2,\mathbf{q}_3)$ the generalized response function in cubic order; see Ref.~\cite{Dornheim_PRR_2021} for a more detailed derivation.

In addition, we note that the generalized quadratic density response function (see Ref.~\cite{Dornheim_JCP_ITCF_2021} for details) is connected to the imaginary-time structure of the system by the relation
\begin{eqnarray}\label{eq:Y_imaginary}
\lefteqn{{Y}(\qv_1,\qv_2)=\frac{1}{2L^3}\int\limits_0^{\beta}d\tau_1\int\limits_0^{\beta}d\tau_2}&&\\\nonumber
&&\times\langle
\tilde n(\qv_1+\qv_2,0)\tilde n(-\qv_1,-\tau_1)\tilde n(-\qv_2,-\tau_2)
\rangle\,.
\end{eqnarray}
Here we define the density operator
\begin{eqnarray}\label{eq:n_tilde}
\tilde n(\mathbf{q},\tau) = \sum_{l=1}^N \textnormal{exp}\left(
-i\mathbf{q}\cdot\hat{\mathbf{r}}_{l,\tau}
\right)
\end{eqnarray}
to be not normalized, and $\mathbf{r}_{l,\tau}$ denotes the position of particle $l$ at an imaginary time $\tau\in[0,\beta]$. Eq.~(\ref{eq:Y_imaginary}) thus directly implies that all quadratic terms of the nonlinear density response (including mode-coupling effects, see below) can be obtained from a single simulation of the unperturbed UEG. Moreover, a similar relation exists for the cubic response function $Z(\mathbf{q}_1,\mathbf{q}_2,\mathbf{q}_3)$, but it is not employed in the present work.

While Eq.~(\ref{eq:Y_imaginary}) is exact, it is highly desirable to have an accurate theory for the density response that can be evaluated without the need for a computationally expensive PIMC simulation. To this end, we follow the considerations from Ref.~\cite{Dornheim_PRR_2021}, and give an (approximate) mean-field expression for the quadratic density response, taking the
form \cite{PhysRevB.37.9268}
\begin{eqnarray}
\label{eq:Y_RPA}
    {Y}(\vec k-\vec q, \vec q)&=& {Y_0}(\vec k-\vec q, \vec q)\\\nonumber & &\times\left[1-v(\vec q)\chi^{(1)}_{0}(\vec q)\right]^{-1}\\\nonumber & &\times\left[1-v(\vec k)\chi_0^{(1)}(\vec k) \right]^{-1}\\\nonumber & &\times\left[1-v(\vec k-\vec q)\chi_0^{(1)}(\vec k-\vec q) \right]^{-1}\,.
\end{eqnarray}

Moreover, the screening terms in the denominator of Eq.~(\ref{eq:Y_RPA}) can be improved by including electronic XC-effects in the form of the effectively static LFC $\overline{G}(\mathbf{k})$, which leads to
\begin{eqnarray}
\label{eq:Y_LFC}
    {Y}(\vec k-\vec q, \vec q) &=& {Y_0}(\vec k-\vec q, \vec q)\\\nonumber & &\times \left[1-v(\vec q)\left(1-\overline{G}(\mathbf{q})\right)\chi^{(1)}_{0}(\vec q)\right]^{-1}\\\nonumber & &\times\left[1-v(\vec k)\left(1-\overline{G}(\mathbf{k})\right)\chi_0^{(1)}(\vec k) \right]^{-1}\\\nonumber & &\times\left[1-v(\vec k-\vec q)\left(1-\overline{G}(\mathbf{k}-\mathbf{q})\right)\right.\\\nonumber
     & &\qquad \qquad \quad \qquad \quad \, \left.\chi_0^{(1)}(\vec k-\vec q) \right]^{-1}\,.
\end{eqnarray}
In practice, we employ the analytical representation of $\overline{G}(\mathbf{k};r_s,\theta)$ from Ref.~\cite{Dornheim_PRB_ESA_2021}.

\section{Results\label{sec:results}}

All PIMC simulation results that are presented in this work have been obtained using a canonical adaption~\cite{mezza} of the worm algorithm by Boninsegni \textit{et al.}~\cite{boninsegni1,boninsegni2}. Further, we employ a primitive factorization of the density matrix, and the convergence with the number of imaginary-time steps $P$ has been carefully checked.

\begin{figure*}\centering
\includegraphics[width=0.475\textwidth]{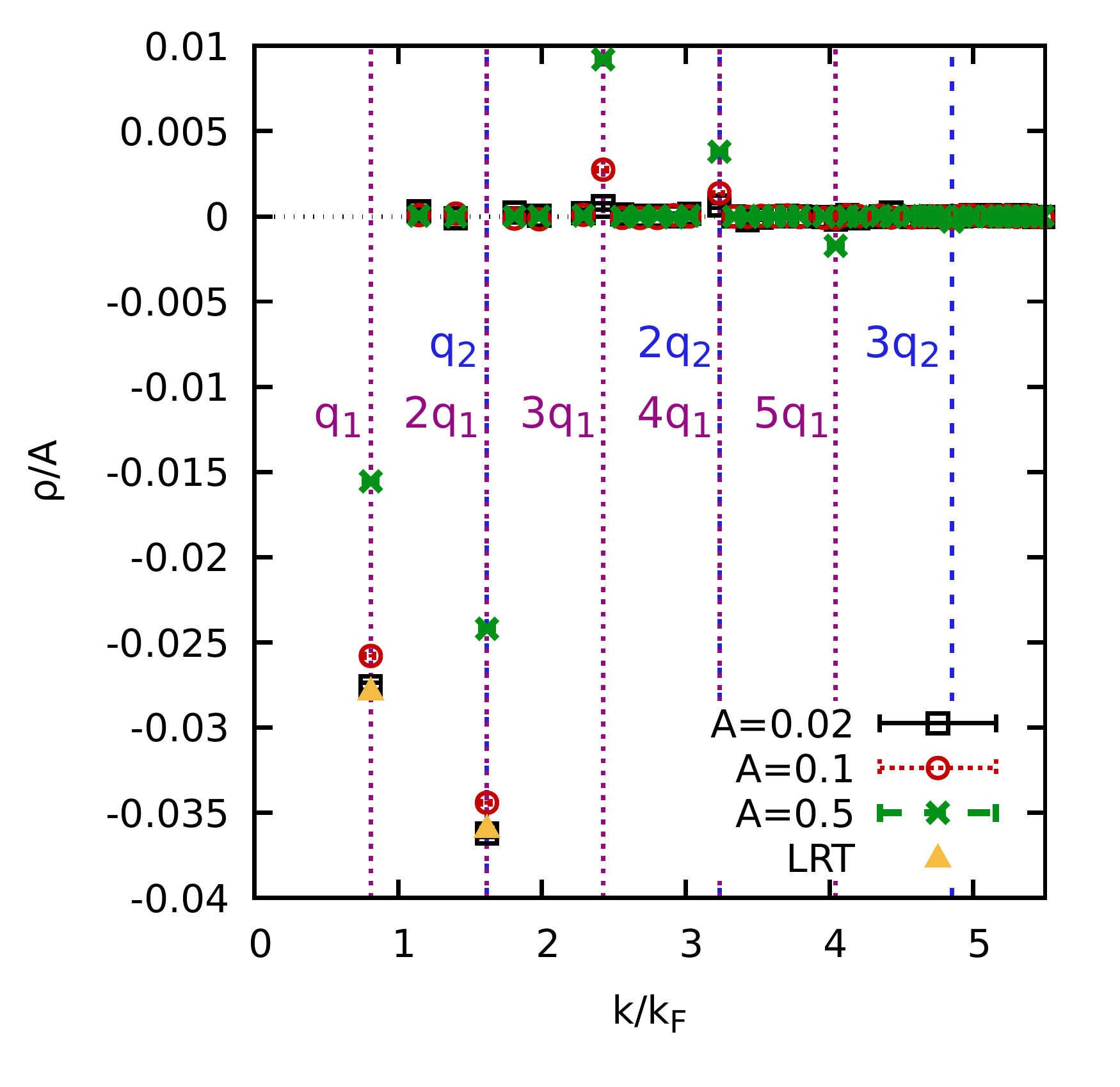}
\includegraphics[width=0.475\textwidth]{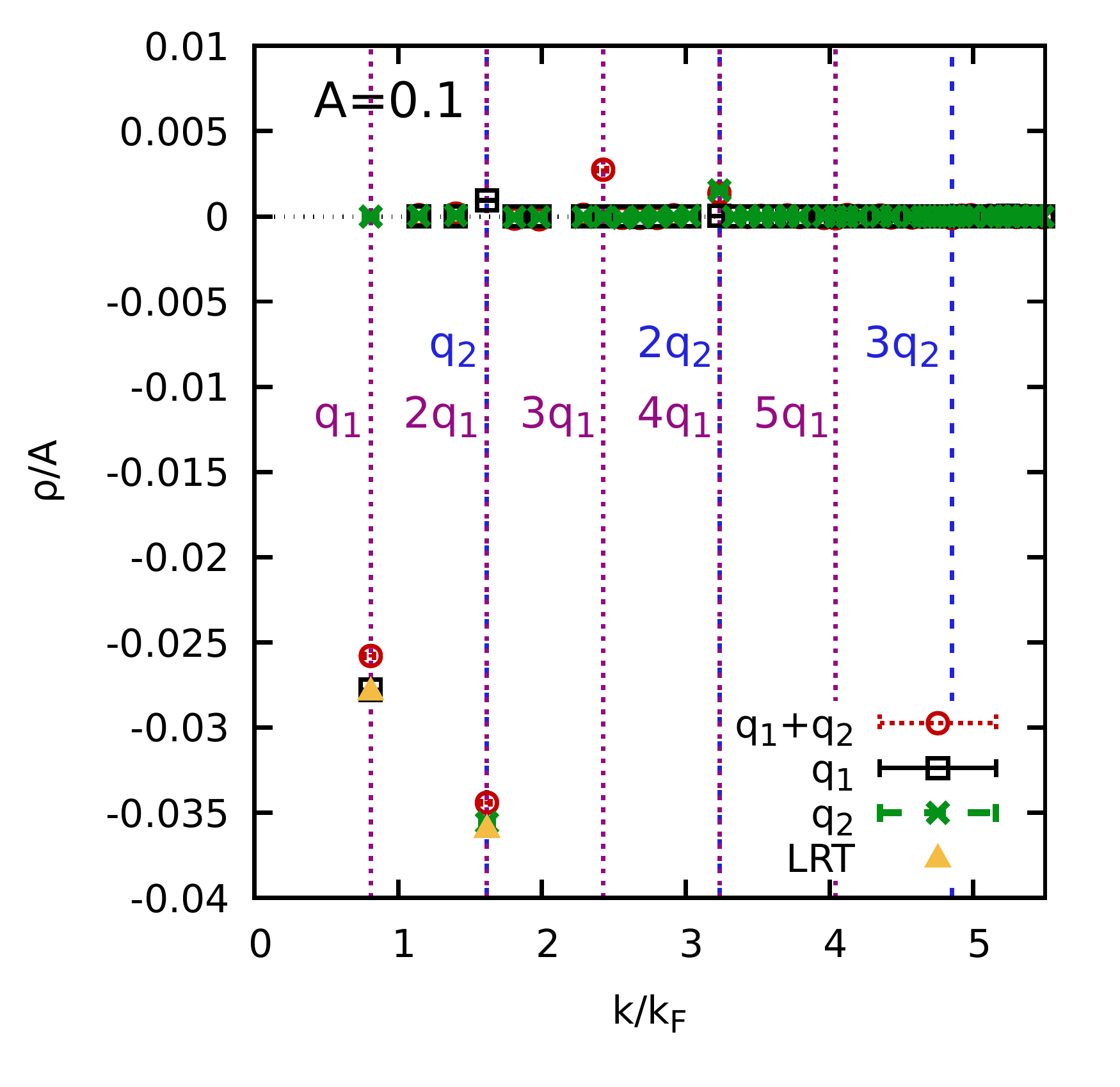}
\caption{\label{fig:spectrum_rs2_theta1}
Wave-number dependence of the density response $\braket{\hat\rho_\mathbf{k}}_{\{\mathbf{q}\},\{A\}}$ for $N_A=2$ perturbations with $A_1=A_2=A$, $\mathbf{q_1}=(1,0,0)^T2\pi/L$, and $\mathbf{q_2}=(2,0,0)^T2\pi/L$ for $N=14$ electrons at $r_s=2$ and $\theta=1$. Left panel: Response spectrum for different perturbation amplitudes, $A=0.02$ (black squares), $A=0.1$ (red circles), and $A=0.5$ (green crosses). Yellow triangles: LRT. Right panel: Response spectrum for $A=0.1$. Red squares: $\braket{\hat\rho_\mathbf{k}}_{\{\mathbf{q}\},\{A\}}$ (same as in the left panel); black squares (green crosses): Response for a single perturbation with $\mathbf{q}=\mathbf{q_1}$ ($\mathbf{q}=\mathbf{q_2}$), taken from Ref.~\cite{Dornheim_PRR_2021}.
}
\end{figure*}

Let us start this investigation by considering the full wave-number dependence of the density response for the case of two harmonic perturbations at $r_s=2$ and $\theta=1$ shown in Fig.~\ref{fig:spectrum_rs2_theta1}. We note that this parameter combination corresponds to a metallic density (e.g. aluminum~\cite{Sperling_PRL_2015}) in the WDM regime, and has frequently been studied in previous investigations~\cite{Dornheim_PRR_2021,dornheim_prl_20,moldabekov2021relevance,Dornheim_JPSJ_2021,Dornheim_CPP_2021,Dornheim_JCP_ITCF_2021, moldabekov2021benchmarking} of the nonlinear electronic density response.
The left panel has been obtained for the case of equal perturbation amplitudes, $A_1=A_2=A$, and for the two wave numbers $\mathbf{q_1}=(1,0,0)^T\ 2\pi/L$, and $\mathbf{q_2}=(2,0,0)^T \ 2\pi/L$. The yellow triangles depict the prediction from LRT, which is given by two independent signals at $\mathbf{q}_1$ and $\mathbf{q}_2$; any interplay between multiple perturbations is completely neglected.

The black squares have been obtained from our PIMC simulations with a comparably small perturbation amplitude $A=0.02$. In this case, LRT is relatively accurate and the density response is close to zero for all $\mathbf{k}\neq\mathbf{q}_i$, although small yet significant signals can be seen at $\mathbf{k}=3\mathbf{q}_1$ and $\mathbf{k}=2\mathbf{q}_2$. The latter effect is substantially increased for $A=0.1$, as can be seen from the red circles in the same plot. Finally, the green crosses have been obtained for the case of a strong perturbation, $A=0.5$, and the depicted spectrum of the density response strongly deviates from LRT and exhibits at least five different signals.

To understand these nontrivial findings, we have to evaluate the different contributions to the mode-coupled density response given in Eq.~(\ref{eq:response}). Firstly, the first line corresponds to the usual LRT contributions at the original wave numbers $\mathbf{q}_i$, which are always present. The second part corresponds to all quadratic contributions, which, in turn, can be further sub-divided into two distinct categories: i) the case $i=j$ gives the quadratic density response at the second harmonic of the respective perturbation. This, too, is not a mode-coupling effect and has been previously reported in Refs.~\cite{Dornheim_PRR_2021,Dornheim_JCP_ITCF_2021}; ii) the case $i\neq j$ entails the mode-coupling signal, and predicts a quadratic density response for $\pm \mathbf{k}=\mathbf{q}_1\pm\mathbf{q}_2$. For the present case, this results in signals at $\mathbf{k}=\mathbf{q}_1$ (which comes in addition to the LRT term) and at $\mathbf{k}=3\mathbf{q}_1$. In fact, the latter constitutes the dominant nonlinear signal at these parameters, which, too, can be easily understood from Eq.~(\ref{eq:response}). Following the notation from Ref.~\cite{Dornheim_PRR_2021}, the density response at the second harmonic of a perturbation at wave number $\mathbf{q}$ is given by the quadratic density response function $\chi^{(2)}(\mathbf{q})=Y(\mathbf{k}-\mathbf{q},\mathbf{q})\delta_{\mathbf{k},2\mathbf{q}}$. The mode-coupling response at $\mathbf{k}=3\mathbf{q}_1$ is instead given by 
\begin{eqnarray}\label{eq:achtung}
\chi^{(2)}_{\mathbf{q}_1+\mathbf{q}_2} &=& Y(\mathbf{k}-\mathbf{q}_1,\mathbf{q}_1)\delta_{\mathbf{k},\mathbf{q}_1+\mathbf{q}_2} \\\nonumber &+& Y(\mathbf{k}-\mathbf{q}_2,\mathbf{q}_2)\delta_{\mathbf{k},\mathbf{q}_1+\mathbf{q}_2}\ ,
\end{eqnarray}
and thus entails two evaluations of the generalized quadratic response function, namely $Y(\mathbf{q}_2,\mathbf{q}_1)$ and $Y(\mathbf{q}_1,\mathbf{q}_2)$.

The last effect from the left panel of Fig.~\ref{fig:spectrum_rs2_theta1} that requires an explanation is the signal at $\mathbf{k}=5\mathbf{q}_1$ that emerges for strong perturbations, $A=0.5$. Evidently, this has to be a mode-coupling effect, as it is not at an integer harmonic of $\mathbf{q}_2$ and would correspond to the fifth harmonic of $\mathbf{q}_1$, which is negligible for such parameters~\cite{Dornheim_PRR_2021}. More specifically, it corresponds to a cubic contribution to Eq.~(\ref{eq:response}) at $\mathbf{k}=2\mathbf{q}_2+\mathbf{q}_1$ that is described by the generalized cubic density response function $Z$.

Let us next consider the right panel of Fig.~\ref{fig:spectrum_rs2_theta1}, where we compare the wave number dependence of the density response from different scenarios. Again, the yellow triangles depict the linear response and have been included as a reference. The red circles correspond to the doubly perturbed simulation for $A=0.1$, i.e., the same as the red circles in the left panel. Furthermore, the black squares and green crosses have been obtained from PIMC simulations with only a single perturbation at $\mathbf{q}_1$ and $\mathbf{q}_2$, respectively, and are taken from Ref.~\cite{Dornheim_PRR_2021}. For the black squares, the signal at the original perturbation can hardly be distinguished from the LRT prediction, whereas the corresponding red circle exhibits a substantially reduced response. Naturally, the latter is a direct consequence of the quadratic mode coupling, as $\mathbf{q}_1=\mathbf{q}_2-\mathbf{q}_1$ in this case. In addition, we find a weak signal at $\mathbf{k}=2\mathbf{q}_1$ for the black circles, which is the incipient quadratic response at the second harmonic. The green crosses exhibit a similar behaviour, although the signal at $\mathbf{k}=\mathbf{q}_2$ is not as strongly reduced for the double perturbation compared to the other points. This makes sense, as the main deviation of the red circles from LRT is given by the aforementioned contribution from the second harmonic of $\mathbf{q}_1$, which is smaller in magnitude than the mode-coupling effect at $\mathbf{k}=\mathbf{q}_1$. 

In addition, we find a signal at the second harmonic of $\mathbf{q}_2$, which is similarly pronounced for both the green crosses and the red circles. Finally, we stress that the signal at $\mathbf{k}=3\mathbf{q}_1$ is only present in the simulation of two external perturbations, as it is expected. We thus briefly summarise the following interim conclusions: mode-coupling i)  typically constitutes the strongest nonlinear effect as it is comprised of multiple contributions at the same wave number $\mathbf{k}$ [cf.~Eq.~(\ref{eq:achtung})] and ii) leads to signals at wave numbers where, otherwise, the response would vanish.

\begin{figure}\centering
\includegraphics[width=0.475\textwidth]{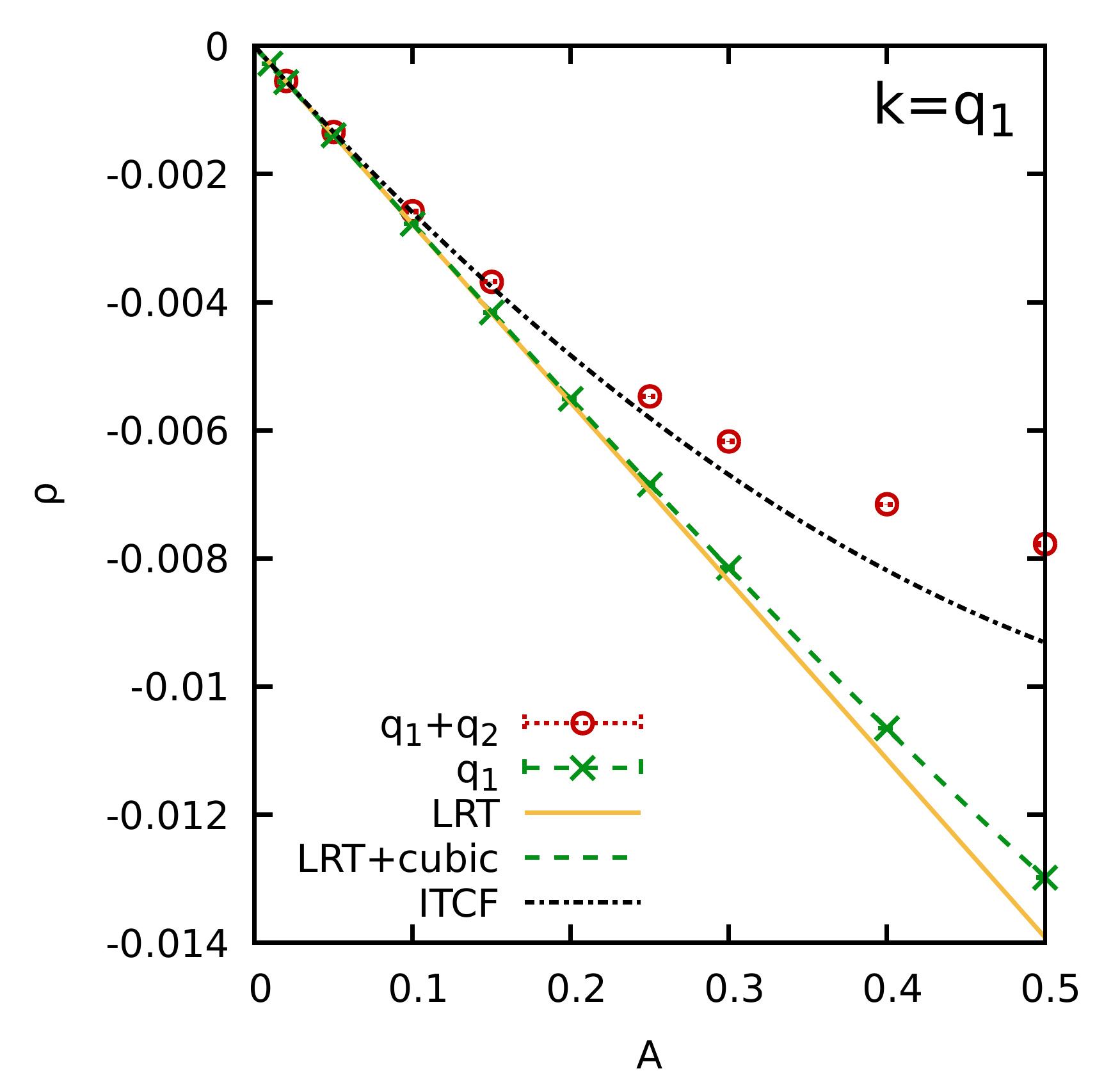}\\\vspace*{-1cm}
\includegraphics[width=0.475\textwidth]{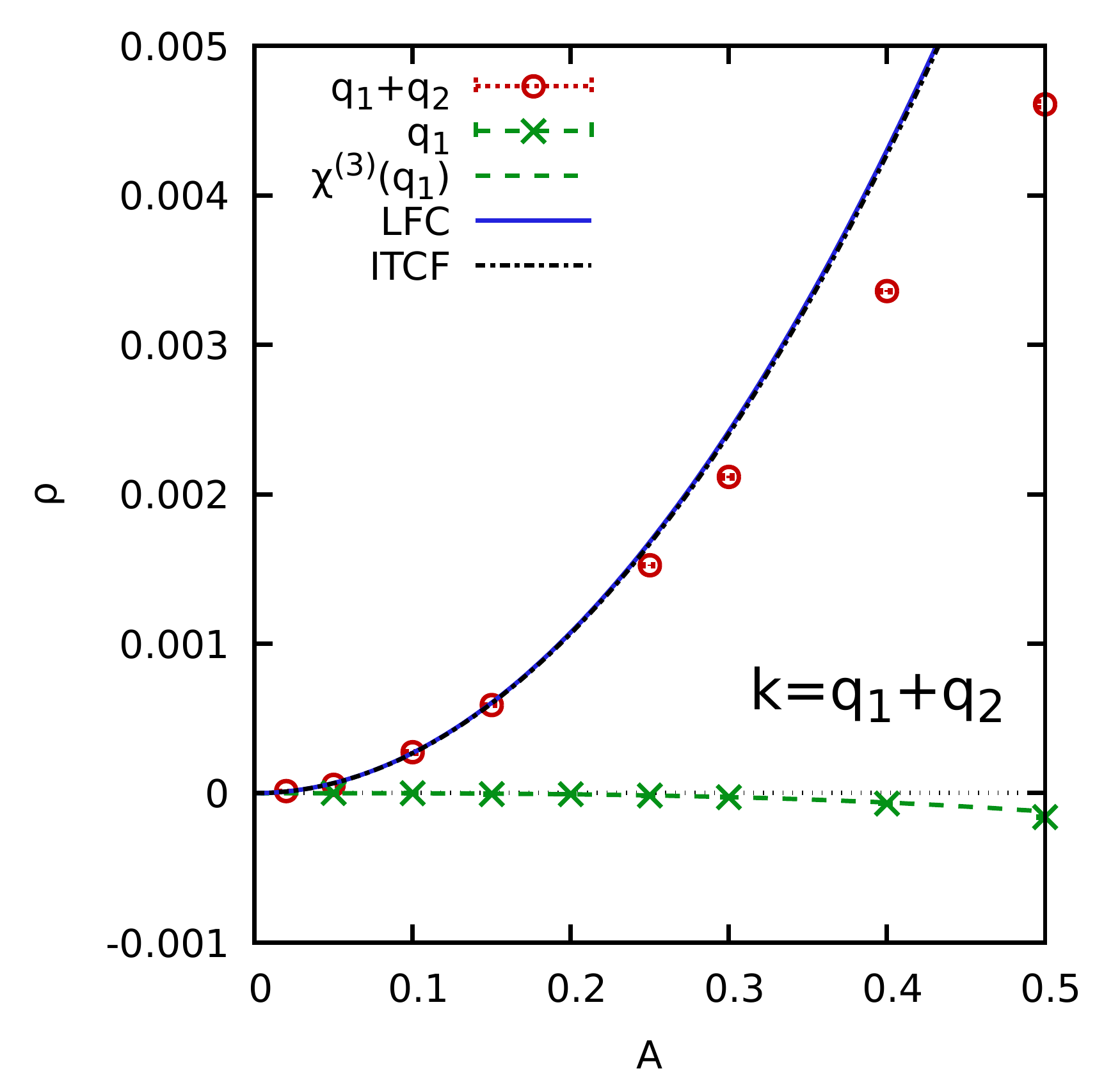}
\caption{\label{fig:deviation_rs2_theta1_qx1}
Dependence of the density response at $\mathbf{k}=\mathbf{q}_1$ (top) and $\mathbf{k}=\mathbf{q}_1+\mathbf{q}_2$ (bottom) on the perturbation amplitude $A$ for the same conditions as in Fig.~\ref{fig:spectrum_rs2_theta1}. Red circles: PIMC results for the density response at $\mathbf{k}$; dash-dotted black: $A\to0$ limit predicted by the ITCF, Eq.~(\ref{eq:Y_imaginary}); green crosses: PIMC results for the density response at $\mathbf{k}$ from a simulation with a single harmonic potential at $\mathbf{q}_1$ taken from Ref.~\cite{Dornheim_PRR_2021}; dashed green: corresponding description based on the cubic density response; solid yellow: LRT; solid blue: LFC-based description of the $A\to0$ limit from Eq.~(\ref{eq:Y_LFC}).
}
\end{figure}

To provide a more quantitative assessment of the effects due to mode-coupling, we investigate the perturbation amplitude dependence of the density response in Fig.~\ref{fig:deviation_rs2_theta1_qx1}. In particular, the top panel corresponds to the case of $\mathbf{k}=\mathbf{q}_1$, i.e., at one of the original perturbations, and the red circles show our new PIMC data that have been obtained for the case of two perturbations for different $A$. As a reference, we also include PIMC data from Ref.~\cite{Dornheim_PRR_2021} that have been obtained for a single perturbation at $\mathbf{q}_1$ as the green crosses.
The solid yellow line shows the prediction from LRT, which is in good agreement to both PIMC data sets for very small $A$. Strikingly, LRT holds for substantially larger $A$ in the case of a single perturbation compared to the more complicated behaviour due to mode-coupling between $\mathbf{q}_1$ and $\mathbf{q}_2$. In fact, the dominant nonlinear effect in the green crosses is given by the cubic density response at the first harmonic, which is depicted by the dashed green line, and is in excellent agreement to the data points. 
In stark contrast, the first nonlinear term in the red circles is quadratic in the perturbation amplitude $A$ and determined by the generalized response function $Y(-\mathbf{q}_1,\mathbf{q}_2)=Y(\mathbf{q}_2,-\mathbf{q}_1)$. The latter can be conveniently estimated by performing a PIMC simulation of the unperturbed electron gas and evaluating Eq.~(\ref{eq:Y_imaginary}); see Ref.~\cite{Dornheim_JCP_ITCF_2021} for a detailed description of this procedure. The final result for both LRT and the thus obtained quadratic mode-coupling term is depicted by the dash-dotted black curve, which is in excellent agreement to the independent data points for $A\leq0.2$. For larger $A$, cubic terms in Eq.~(\ref{eq:response}) start to become noticeable, as it is expected.

The bottom panel of Fig.~\ref{fig:deviation_rs2_theta1_qx1} shows the same information, but for $\mathbf{k}=\mathbf{q}_1+\mathbf{q}_2$. In this case, the green crosses exhibit a very weak response that is well described by the cubic response of the third harmonic of $\mathbf{q}_1$, i.e., the dashed green curve. In stark contrast, the red circles are larger by an order of magnitude and exhibit a parabolic behaviour [cf.~Eq.~(\ref{eq:achtung})] in leading order that is again well described by the theoretical curve that we obtain by evaluating Eq.~(\ref{eq:Y_imaginary}). In addition, the solid blue curve has been obtained by evaluating Eq.~(\ref{eq:Y_LFC}) using as input the static local field correction by Dornheim \emph{et al.}~\cite{Dornheim_PRL_2020_ESA}. Evidently, the resulting parabola is in excellent agreement to the exact $A\to0$ limit predicted by the ITCF formalism. This, in turn, further corroborates the high value of the LFC for the description of nonlinear effects that has been reported in a previous work~\cite{Dornheim_PRR_2021}.


\begin{figure}\centering
\includegraphics[width=0.475\textwidth]{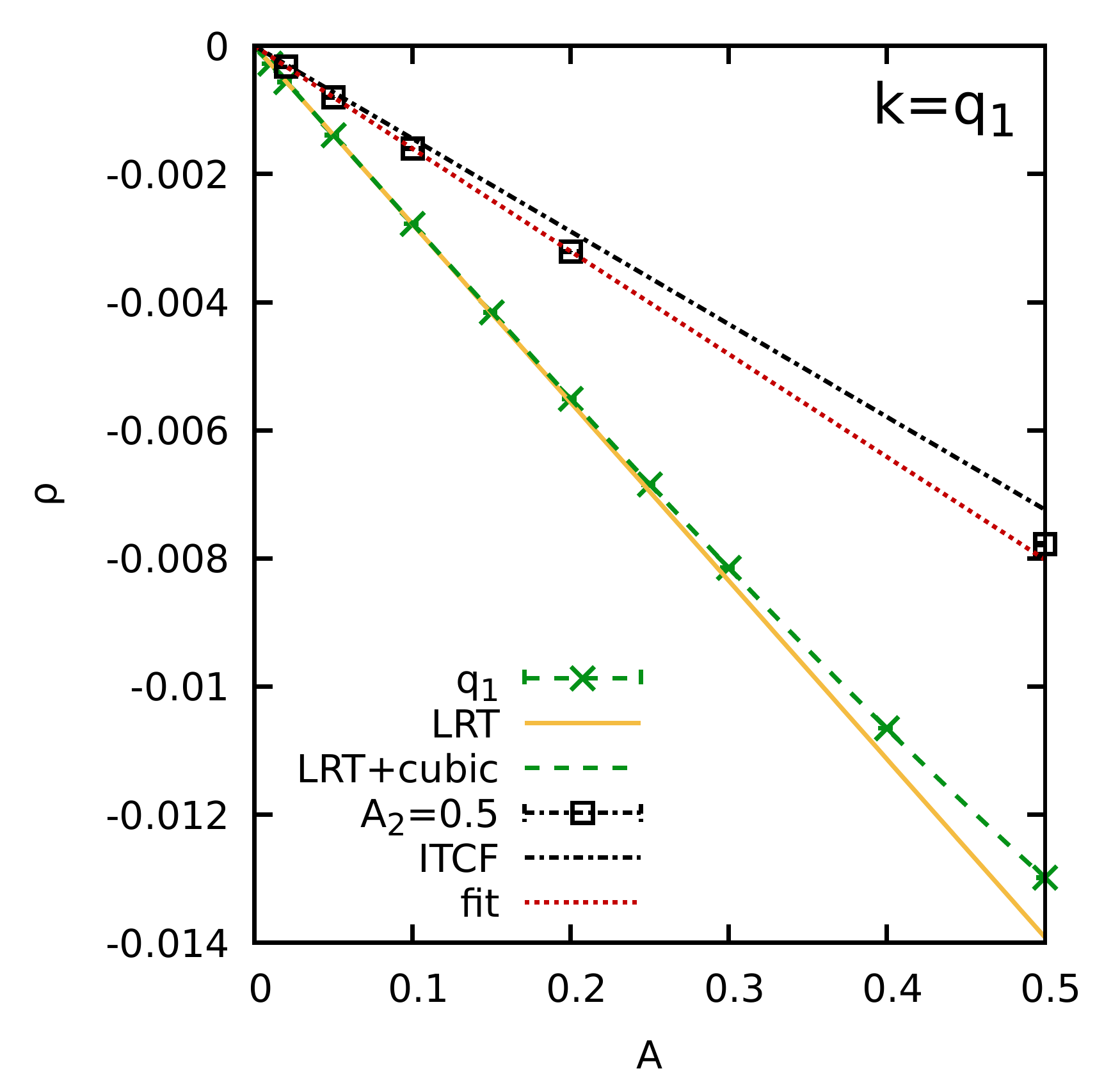}
\caption{\label{fig:Double}
Perturbation strength ($A_1$) dependence of the density response at $\mathbf{q_1}=(1,0,0)^T2\pi/L$ for $r_s=2$ and $\theta=1$. A second perturbation at $\mathbf{q_2}=(2,0,0)^T2\pi/L$ is applied with $A_2=0.5$ being fixed. Black squares: PIMC data for the density response at $\mathbf{k}=\mathbf{q}_1$; dash-dotted black: ITCF prediction, Eq.~(\ref{eq:Y_imaginary}); dotted red: linear fit for $A\leq0.2$; green crosses: PIMC results for a single external perturbation at $\mathbf{q}_1$ taken from Ref.~\cite{Dornheim_PRR_2021}; solid yellow: LRT prediction for the latter case; dashed green: corresponding LRT+cubic response.
}
\end{figure}

A further interesting topic is the behaviour of the density response when the perturbation amplitudes of the two perturbations are not equal. Such a case is investigated in Fig.~\ref{fig:Double}, where we consider the same wave vectors $\mathbf{q}_1$ and $\mathbf{q}_2$ as before. More specifically, we show the density response at $\mathbf{k}=\mathbf{q}_1$, and the $x$-axis shows the perturbation amplitude $A_1=A$; the perturbation amplitude $A_2=0.5$ is being kept constant. The resulting data are shown as the black squares, and, approximately, exhibit a linear behaviour with $A$. At the same time, the pre-factor substantially differs from the LRT prediction (solid yellow), and from the simulation results obtained for only a single perturbation (green crosses). 
In other words, we find an apparently linear effect that is not described by LRT.

Yet, this seeming contradiction is resolved by the second line in Eq.~(\ref{eq:response}), which predicts a mode-coupling contribution at $\mathbf{k}=\mathbf{q}_1$ that is proportional to $A_1\times A_2$. Since $A_2$ is being kept constant for the results shown in Fig.~\ref{fig:Double}, the \emph{quadratic} mode-coupling term manifests with a linear dependence on $A_1=A$, which resolves this apparent conundrum. The corresponding estimation combining both LRT and the quadratic term that has been obtained from the generalized ITCF [Eq.~(\ref{eq:Y_imaginary})] is depicted as the dash-dotted black line in Fig.~\ref{fig:Double}, and qualitatively captures the correct trend. The comparably small residual differences between data and theory are due to the fact that at $A_2=0.5$, even the quadratic description is not fully sufficient; see Fig.~\ref{fig:deviation_rs2_theta1_qx1} above. Finally, the dotted red line has been obtained from a linear fit to the PIMC data for $A\leq0.2$, and, therefore, depicts the correct nonlinear result for the linear response in this case.
We stress that this finding might be of considerable practical relevance, as it directly implies that the presence of a second external perturbation with a large amplitude leads to a break-down of LRT for the first perturbation even in the limit of $A\to 0$. For example, an external ionic potential might be expanded into such harmonics, and the corresponding coefficients might be substantial.

\begin{figure}\centering
\includegraphics[width=0.475\textwidth]{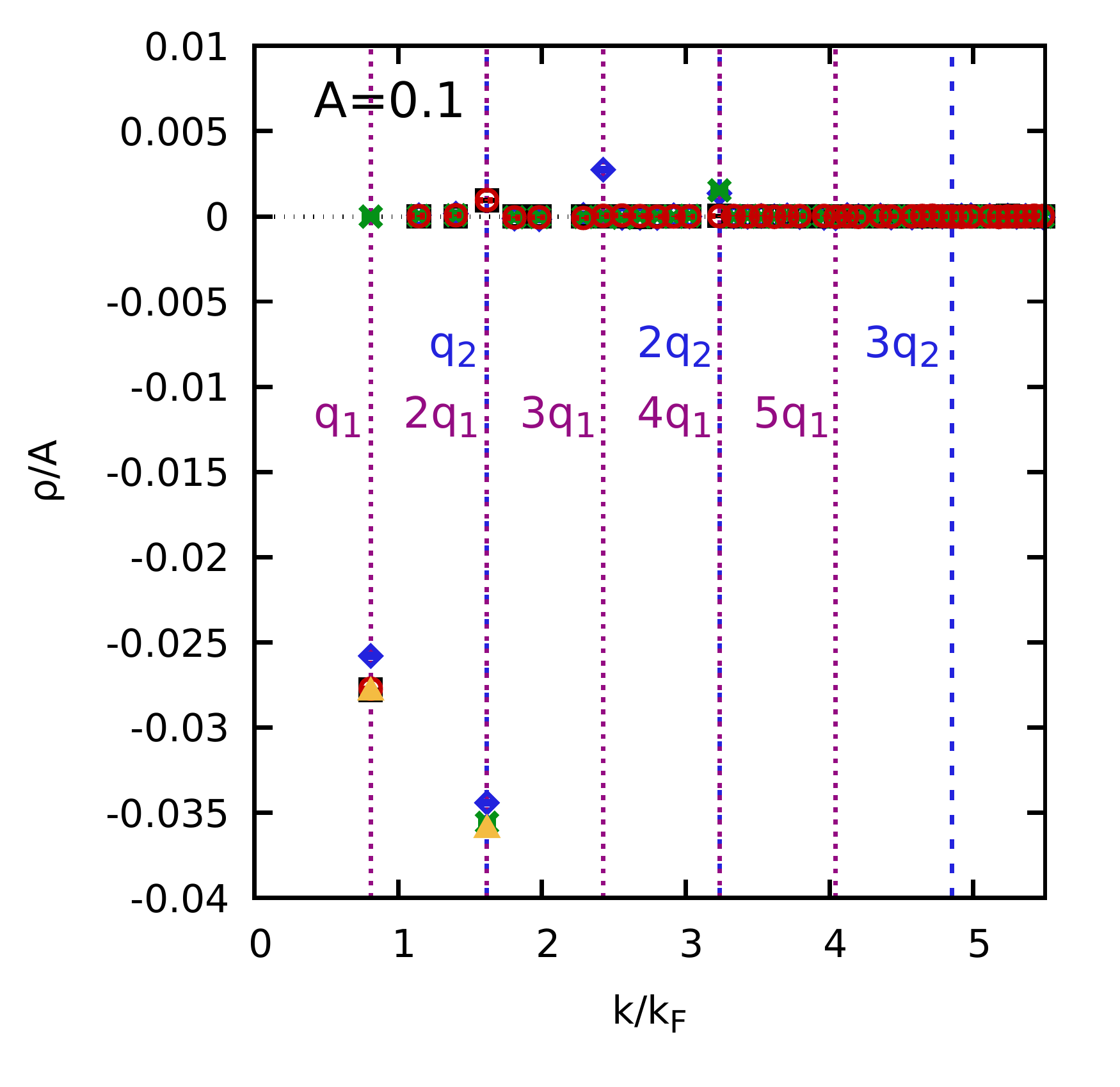}
\caption{\label{fig:direction}
Wave-number dependence of the density response $\braket{\hat\rho_\mathbf{k}}_{\{\mathbf{q}\},\{A\}}$ at $r_s=2$ and $\theta=1$ for $A=0.1$. Black squares: Single perturbation along $\mathbf{q_1}=(1,0,0)^T2\pi/L$; Green crosses: Single perturbation along $\mathbf{q_2}=(2,0,0)^T2\pi/L$ . Blue diamonds: two perturbations along $\mathbf{q_1}$ and $\mathbf{q_2}$. Red circles: two perturbations along $\mathbf{q_1}$ and $\mathbf{q_3}=(0,2,0)^T2\pi/L$. Yellow triangles: LRT.
}
\end{figure}

Let us next consider the impact of the direction of the external perturbation. To this end, we investigate the wave number dependence of the density response again for $r_s=2$ and $\theta=1$ in Fig.~\ref{fig:direction}. Specifically, the black squares and green crosses have been obtained for the case of single perturbations along $\mathbf{q}_1$ and $\mathbf{q}_2$, respectively, with $A=0.1$. In addition, the blue diamonds show the signal for the case of a double perturbation at both $\mathbf{q}_1$ and $\mathbf{q}_2$ with $A_1=A_2=A$. Finally, the red circles have been obtained from a new simulation with two perturbations along $\mathbf{q}_1$ and $\mathbf{q}_3=(0,2,0)^T\ 2\pi/L$. We note that the $k$-vectors shown in Fig.~\ref{fig:direction} have been selected along the $x$-direction insofar as this is possible. Therefore, the red circles are indistinguishable from the black squares as no mode-coupling affects the density response in this direction.

\begin{figure}\centering
\includegraphics[width=0.475\textwidth]{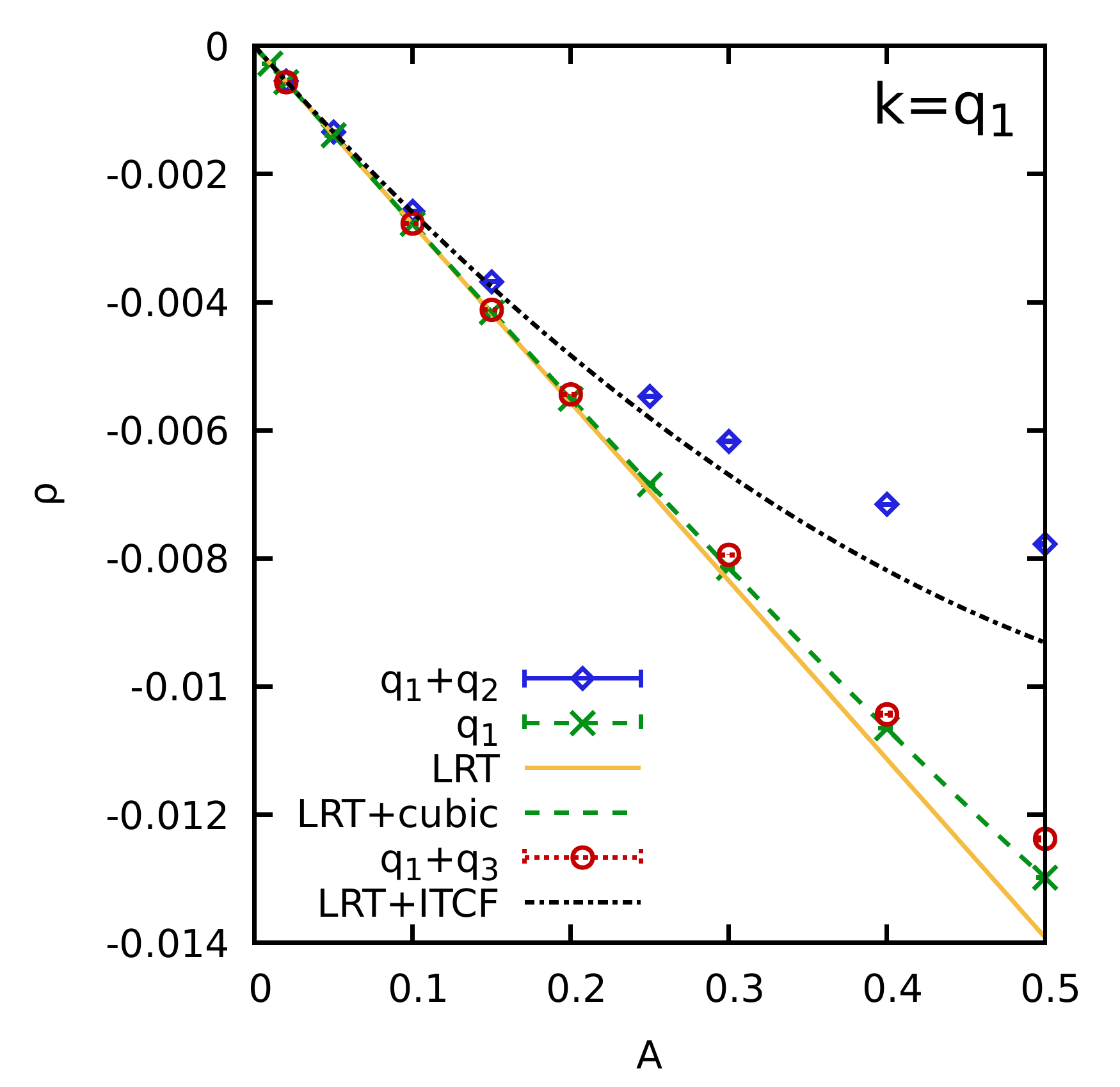}
\caption{\label{fig:direction_A}
Density response at $\mathbf{q_1}=(1,0,0)^T2\pi/L$ for the same conditions as in Fig.~\ref{fig:direction}.
}
\end{figure} 

This notion is further validated in Fig.~\ref{fig:direction_A}, where we show the $A$-dependence of the density response at $\mathbf{k}=\mathbf{q}_1$. The blue diamonds show the PIMC data for the simulation with two perturbations along $\mathbf{q}_1$ and $\mathbf{q}_2$, which are well described by LRT and the quadratic mode-coupling response (dash-dotted black curve) as it has been explained before. The green crosses are the simulation results for only a single perturbation at $\mathbf{q}_1$, which follow the combination of LRT and the cubic response at the first harmonic. 
Finally, the red circles have been obtained from the simulations with perturbations at $\mathbf{q}_1$ and $\mathbf{q}_3$. Evidently, the response at $\mathbf{k}=\mathbf{q}_1$ is hardly affected by the presence of the second perturbation along another direction, and the circles closely follow the green crosses for small to medium perturbations. For completeness, we mention that the small deviations at large perturbation amplitudes $A$ are a consequence of cubic mode-coupling terms, cf.~Eq.~(\ref{eq:response}).

\begin{figure*}\centering
\includegraphics[width=0.71\textwidth]{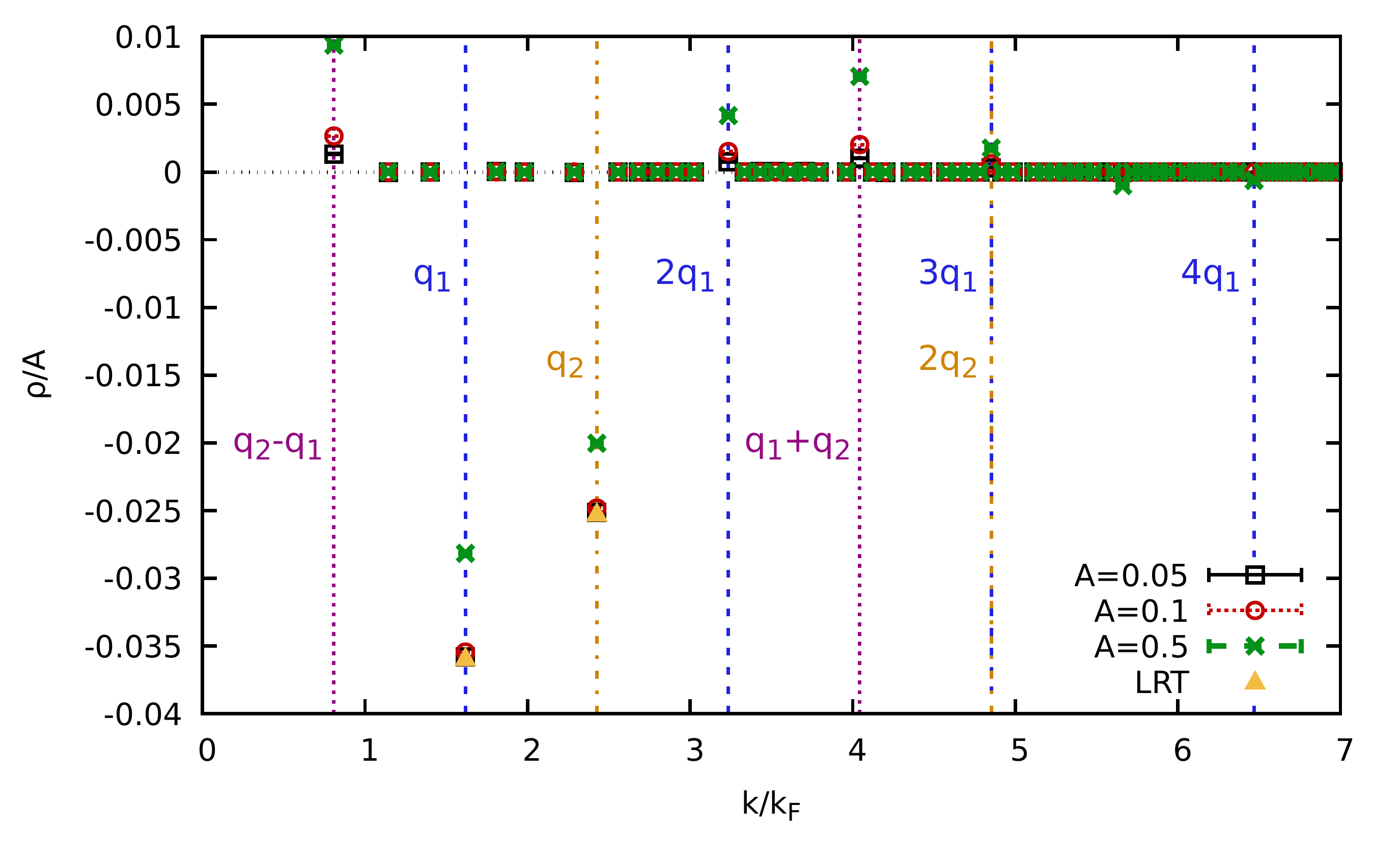}
\caption{\label{fig:prime_spectrum}
Wave-number dependence of the density response $\braket{\hat\rho_\mathbf{k}}_{\{\mathbf{q}\},\{A\}}$ at $r_s=2$ and $\theta=1$ for two perturbations with $\mathbf{q_1}=(2,0,0)^T2\pi/L$ and $\mathbf{q_2}=(3,0,0)^T2\pi/L$, and $A_1=A_2=A$.
Black squares: $A=0.05$; Red circles: $A=0.1$; Green crosses: $A=0.5$; Yellow triangles: LRT.
}
\end{figure*}

Let us conclude our study of mode-coupling effects in the density response of the warm dense UEG with a more complicated combination of perturbation wave vectors. This situation is investigated in Fig.~\ref{fig:prime_spectrum}, where we consider the case of $\mathbf{q}_1=(2,0,0)^T\ 2\pi/L$ and $\mathbf{q}_2=(3,0,0)^T\ 2\pi/L$, again with $A_1=A_2=A$. In particular, the $\mathbf{q}L/2\pi$ are given by prime numbers, so that no quadratic mode-coupling contribution coincides with the original wave numbers or their integer harmonics. As usual, the yellow triangles depict the prediction from LRT, and the black squares, red circles, and green crosses show PIMC results for $A=0.05$, $A=0.1$, and $A=0.5$, respectively.

Evidently, this choice of the $\mathbf{q}_i$ results in a number of signals at different wave vectors $\mathbf{k}$ even in the case of a rather weak perturbation, $A=0.05$. More specifically, there appear four quadratic signals (the second harmonics of $\mathbf{q}_1$ and $\mathbf{q}_2$, and the mode-coupling signals at $\mathbf{k}=\mathbf{q}_2\pm\mathbf{q}_1$) in addition to the two main peaks at the original wave vectors of the external perturbation.
In the regime of large perturbations, $A=0.5$, the situation becomes even more complicated as additional signals occur due to cubic effects.
We thus conclude that the density response to a superposition of multiple external perturbations becomes dramatically more complicated beyond the LRT regime.

\section{Summary and Discussion\label{sec:summary}}

In summary, we have analyzed nonlinear mode-coupling effects in the density response of an electron gas in the WDM regime. This has been achieved on the basis of extensive new \emph{ab initio} PIMC results that have been obtained by applying multiple external harmonic perturbations at the same time. First and foremost, we note that mode-coupling effects, while being absent in LRT, constitute the dominant nonlinear effect for weak to moderate perturbation amplitudes. In addition, the presence of a strong, constant perturbation at $\mathbf{q_2}$ leads to a substantial modification of the linear response at a different wave vector $\mathbf{q}_1$ even in the limit of $A\to0$, which has considerable implications for a non-uniform electron gas in an external potential, for example due to positively charged ions. 

From a theoretical perspective, we have shown that mode-coupling effects can be estimated numerically exactly from a PIMC simulation of the unperturbed UEG by computing the generalized imaginary-time correlation functions that have recently been introduced by Dornheim \emph{et al.}~\cite{Dornheim_JCP_ITCF_2021}. This leads to a substantial reduction of the computational effort, as, in principle, the full information about mode-coupling between all wave vectors can be obtained from a single simulation.

In addition, we have extended our earlier analytical theory~\cite{Dornheim_PRR_2021} for the nonlinear density response in terms of the static LFC~\cite{dornheim_ML,Dornheim_PRL_2020_ESA,Dornheim_PRB_2021}, and find excellent agreement between theory and simulations with negligible computational cost.

There exists a large variety of nonlinear response phenomena in WDM. An interesting question is to inquire about the physical origin of the nonlinearity. For example, the Coulomb interaction between the electrons automatically gives rise to nonlinearities: the $1/r$ distance dependence of the Coulomb potential transforms a harmonic perturbation of any particle into a nonlinear response of its neighbors. A familiar example is the generation of high harmonics of a laser field in a gas~\cite{Kuh}. Therefore, our results will be particularly important for situations where the plasma is strongly correlated.
While we have studied only moderately coupled warm dense matter ($r_s=2$), the consequences will be even more significant when $r_s$ is increased. At strong correlations a number of additional nonlinear effects should be expected. For example, modes propagating in different directions may couple if the isotropy of a nonideal system is broken, e.g. in the presence of a magnetic field. Such an effect is known from strongly couple classical plasmas where it leads to nontrivial coupling of transport processes in different directions, including diffusion~\cite{PhysRevLett.107.135003} and heat transport~\cite{PhysRevE.92.063105}. Similar nonlinear effects that are mediated by strong Coulomb interaction should also be expected in WDM.

We expect that our new results will open up many avenues for future research in the field of WDM theory and beyond. First and foremost, we mention the potential utility of nonlinear effects in the electronic density response as a method of diagnostics in experiments. Specifically, the interpretation of current X-ray Thomson scattering (XRTS) experiments is solely based on LRT~\cite{siegfried_review}, and the inference of important plasma parameters such as the electronic temperature $T_e$ is notoriously difficult~\cite{kraus_xrts}. In this context, Moldabekov \emph{et al.}~\cite{moldabekov2021thermal} have recently suggested that the application of an external perturbation to generate inhomogeneous WDM samples on purpose leads to a modified experimental signal, that more strongly depends on $T_e$. Naturally, the accurate theoretical description in this case will have to take into account the mode-coupling between this external potential and the XRTS probe, and the current LFC-based theory is uniquely suited for this endeavour.
A second potential application of our work is given by the possibility to use the nonlinear density response as a probe for three- and even four-body correlation functions known from many-body theory~\cite{Dornheim_JPSJ_2021}, which might give unprecedented insights into the physical mechanisms of WDM.
Finally, we mention that a general theory of the nonlinear electronic density response can be directly incorporated into many theoretical approached that have hitherto, often by necessity, been limited to LRT. Prominent examples include the construction of effectively screened ionic potentials~\cite{zhandos1,zhandos2, zhandos_cpp17} and the electronic stopping power~\cite{moldabekov_pre_20}. Specifically, going beyond LRT is a crucial step for understanding and exploring the effective ion-ion attraction in the media with strongly correlated electrons \cite{Bonev, Gravel}, which remains to be an unsolved problem \cite{Schoof_PRE, zhandos_cpp21}.
Additionally, the effect of electronic strong correlations on a projectile energy dissipation (stopping power) is one of the long standing problems in WDM \cite{GRABOWSKI2020100905, PhysRevLett.122.015002}, which can be addressed using the non-linear response theory \cite{PhysRevB.37.9268}.

$ $

$ $

\section*{Acknowledgments}
This work was partly funded by the Center of Advanced Systems Understanding (CASUS) which is financed by Germany's Federal Ministry of Education and Research (BMBF) and by the Saxon Ministry for Science, Culture and Tourism (SMWK) with tax funds on the basis of the budget approved by the Saxon State Parliament, and by the Deutsche Forschungsgemeinschaft (DFG) via project BO1366/15.
The PIMC calculations were carried out at the Norddeutscher Verbund f\"ur Hoch- und H\"ochstleistungsrechnen (HLRN) under grant shp00026, on a Bull Cluster at the Center for Information Services and High Performance Computing (ZIH) at Technische Universit\"at Dresden,
on the cluster \emph{hemera} at Helmholtz-Zentrum Dresden-Rossendorf (HZDR), and at the computing center (Rechenzentrum) of Kiel university.

\bibliography{bibliography,mb-ref}
\end{document}